# Crystal structure, properties and pressure-induced insulator-metal transition in layered kagome chalcogenides

Hong Du[1], Yu Zheng[1], Cuiying Pei[2], Chi-Ming Yim[1,3], Yanpeng Qi[2,4,5], Ruidan Zhong[1,3*]

1. Tsung-Dao Lee Institute, Shanghai Jiao Tong University, Shanghai 201210, China
2. School of Physical Science and Technology, ShanghaiTech University, Shanghai 201210, China
3. School of Physics and Astronomy, Shanghai Jiao Tong University, Shanghai 200240, China
4. ShanghaiTech Laboratory for Topological Physics, ShanghaiTech University, Shanghai 201210, China
5. Shanghai Key Laboratory of High-resolution Electron Microscopy, ShanghaiTech University, Shanghai 201210, China

*E-mail: rzhong@sjtu.edu.cn

**Abstract**

Layered materials with kagome lattice have attracted a lot of attention due to the presence of nontrivial topological bands and correlated electronic states with tunability. In this work, we investigate a unique van der Waals (vdW) material system, $A_2M_3X_4$ ($A$ = K, Rb, Cs; $M$ = Ni, Pd; $X$ = S, Se), where transition metal kagome lattices, chalcogen honeycomb lattices and alkali metal triangular lattices coexist simultaneously. A notable feature of this material is that each Ni/Pd atom is positioned in the center of four chalcogen atoms, forming a local square-planar environment. This crystal field environment results in a low spin state $S$ = 0 of $Ni^{2+}/Pd^{2+}$. A systematic study of the crystal growth, crystal structure, magnetic and transport properties of two representative compounds, $Rb_2Ni_3S_4$ and $Cs_2Ni_3Se_4$, has been carried out on powder and single crystal samples. Both compounds exhibit nonmagnetic $p$-type semiconducting behavior, closely related to the particular chemical environment of $Ni^{2+}$ ions and the alkali metal intercalated vdW structure. Additionally, $Cs_2Ni_3Se_4$ undergoes an insulator-metal transition (IMT) in transport measurements under pressure up to 87.1 GPa without any structural phase transition, while $Rb_2Ni_3S_4$ shows the tendency to be metalized.

## 1. Introduction

Kagome lattice, composed of two-dimensional (2D) corner-sharing triangles, has been a keen research topic in the context of condensed matter physics for the past decades. The particular lattice geometry gives rise to an electronic structure naturally hosting Dirac cones, van Hove singularities, and a completely flat band[1,2]. The presence of nontrivial band topology and strongly correlated electrons can lead to diverse topological and many-body phenomena, such as quantum spin liquid[3], charge density wave[4,5], Dirac/Weyl fermions[6–8], anomalous Hall effect[9], and unconventional superconductivity[4,5,10]. With such rich and intriguing quantum phases, kagome material stands out as a fertile platform to study the interplay between lattice geometry, band topology, magnetism and correlation degrees of freedom. Over the past years, various materials with transition-metal kagome lattice have been extensively investigated, including copper-based kagome $ZnCu_3(OH)Cl_2$[11], manganese-based kagome $REMn_6Sn_6$ ($RE$ = rare-earth elements)[12,13], cobalt-based kagome $Co_3Sn_2S_2$ and $CoSn$[6,14,15], iron-based kagome $Fe_3Sn_2$, $FeSn$ and $FeGe$[16–21], yet there still are less noticed palladium-based kagome $Pd_3P_2S_8$[22,23] and chromium-based kagome $YCr_6Ge_6$[24]. Recently, the layered titanium/vanadium-based kagome metal $AM_3X_5$ ($A$ = K, Rb, Cs; $M$ = Ti, V; $X$ = Sb, Bi) has attracted lots of attention[25–38]. Many exotic properties, such as chiral charge density wave[28,29], intertwined orders[30,36–38], and electronic nematicity[31] were proposed. Even though various kagome systems have already been investigated as mentioned, nickel-based kagome materials have much less material realization.

Crystal structure and symmetry play a critical role in determining physical properties, and searching for structures with interesting motifs is a basic strategy for the exploration of novel material systems. In this work, we focus on nickel/palladium chalcogenides with the general composition $A_2M_3X_4$ (A = K, Rb, Cs; $M$ = Ni, Pd; $X$ = S, Se). The material family is of great interest as it combines transition metal kagome lattices, chalcogen honeycomb lattices and alkali metal triangular lattices in a vdW layered 2D crystal structure, sharing structural similarities with the $RE$Mn$_6$Sn$_6$ and $AM_3X_5$ families. Bronger *et al.*, the first to synthesize this material, reported and discussed the observation of temperature-independent magnetism in both nickel and palladium compounds[39]. Further evidence from magnetic measurements on K$_2$Ni$_3$S$_4$ indicated that the Ni$^{2+}$ in the local square-planar crystal field adopts a low spin ($S$ = 0) configuration[40], and Raman spectra demonstrated lattice stability without structural transition down to 4 K[41]. However, the suggested nonmagnetic ground state is contradicted by subsequent studies that observed weak ferromagnetism[42] or paramagnetism[43]. Meanwhile, band-structure calculations revealed that Rb$_2$Ni$_3$S$_4$ is a moderately correlated band insulator with a narrow gap size of 0.66 eV, and photoemission spectroscopy confirmed the presence of a nearly dispersionless band below the Fermi surface[44]. Considering the inevitable chemical disorder in the material, further electronic transport characterizations are necessary to illuminate its effect on the conductivity of such a narrow-gap semiconductor. Together with the controversies in the magnetic ground state, it is thus imperative to carefully reexamine the magnetic and electrical transport properties of $A_2M_3X_4$. In addition, Rb$_2$Pd$_3$Se$_4$ has recently been reported to be superconducting under high pressure[45], suggesting that this material system might host unconventional superconductivity and nontrivial band topology.

In light of the above, $A_2M_3X_4$ (A = K, Rb, Cs; $M$ = Ni, Pd; $X$ = S, Se) endowed with transition-metal-based kagome lattices appears to be a highly intriguing material system with substantial research potential. In this work, we have grown single crystal samples, reinvestigated the outstanding features in the crystal structure and studied their corresponding magnetic and electrical properties on two representative compounds, Rb$_2$Ni$_3$S$_4$ and Cs$_2$Ni$_3$Se$_4$. We find that both samples can be considered as nonmagnetic, hole-doped semiconductors. A kink in longitudinal resistivity has been observed around 260 K in Rb$_2$Ni$_3$S$_4$, suggesting a possible phase transition. In addition, Cs$_2$Ni$_3$Se$_4$ shows IMT under pressure up to 87.1 GPa due to a possible bandgap closure under compression, while Rb$_2$Ni$_3$S$_4$ shows the tendency to be metalized.

## 2. Experimental methods

Single crystals of $A_2$Ni$_3$X$_4$ were prepared by fusion reaction[39,46] under a flowing Ar atmosphere for sulfide or an Ar/H$_2$ (H$_2$-8%) mixture for selenides (see Supplementary Information for details). Mixtures of pre-dried Rb$_2$CO$_3$ or Cs$_2$CO$_3$ (99.994%), Ni (99.9999%), and S or Se (>99.99%) in a molar ratio of 1: 5: 12 was placed into alumina crucibles as the starting materials. The crucibles were then heated at a rate of 9 °C min$^{-1}$ to 1000 °C and held at this temperature for 0.5-3 h, and then furnace-cooled to room temperature. The as-grown single crystals with layered morphology that can be obtained by mechanical exfoliation from the surface of the reaction products, are shiny thin foil-like (shown in Figure 2(b,c)) with a typical size of $10 \times 10 \times 0.15$ mm$^3$. The crystals easily lost their luster in moist air, so we handled them in glovebox. Polycrystalline Rb$_2$Ni$_3$S$_4$ samples were synthesized via a conventional solid-state reaction method. Stoichiometric amounts of high-purity Rb (99.75%), Ni (99.9999%), and Se (99.9999%) elements were loaded in a quartz ampoule and sealed under vacuum. Then the quartz ampoule was heated up to 850 °C at a rate of 3 °C min$^{-1}$ and dwelled there for 7 hours, followed by a slow cooling to room temperature.

The structural characterization was performed by Powder X-ray diffraction (PXRD), using a Bruker D8 Advance Eco powder diffractometer equipped with (Cu-$K\alpha$ radiation) ($\lambda$ = 1.5418 Å). Rietveld refinements[47] were carried out on the powder diffraction data using the TOPAS v.6.0 software. The field-dependent magnetization and temperature-dependent *dc* magnetic susceptibility between 2 and 300 K were measured on single crystal samples with a total mass of ~ 1 mg using a Physical Property Measurement System (PPMS, Quantum Design) equipped with a vibrating sample magnetometer option. GE varnish was used to attach sample pieces to a standard sample rod. Diamagnetic signals of GE varnish and background were subtracted from the magnetization raw data to obtain the pure sample signals. Resistivity and Hall effect measurements at ambient pressure were carried out using standard four-probe method. The electrical contacts were made with Au-wires and silver paint.

High pressures were generated with a diamond anvil cell (DAC), as described elsewhere[48]. An in situ high-pressure Raman spectroscopy was performed using a Raman spectrometer (Renishaw inVia, UK) with a laser excitation wavelength of 532 nm and low-wavenumber filter. A symmetric DAC with anvil culet sizes of 200 μm was used, with silicon oil as the PTM. The high-pressure electrical resistivity measurements were performed with the van der Pauw four-probe method. A nonmagnetic DAC with anvil culet size of 200 μm was used. A cubic BN/epoxy mixture was inserted between BeCu gasket and Pt electrode as an insulator layer. No pressure transmitting medium was used. The pressure was calibrated by ruby luminescence method at room temperature. Electrical measurements were taken on a customary cryogenic setup.

## 3. Results and discussion

$Rb_2Ni_3S_4$ (or $Cs_2Ni_3Se_4$) crystallizes in the orthorhombic space group *Fmmm* (No. 69). It is stacked with nickel chalcogenides layers intercalated with alkali metal cations, as shown in Figure 1(a). The layered nickel chalcogenide is of particular interest given that the $Ni^{2+}$ ions form a kagome lattice in the *ac* plane. Each Ni atom forms a square planar structure with its adjacent four chalcogenide atoms. These edge-sharing NiS/NiSe planes further intersect at 120° with each other, leading to the formation of a honeycomb prism, as shown in Figure 1(b). Those layers are further stacked along *b*-axis through vdW interactions and separated by triangular layers of alkali metal atoms, which lie exactly above and below the honeycomb center. As shown in Figure 1(c), this material is a perfect realization of the coexisting kagome, honeycomb and triangular motifs. The similar stacking order has been reported in the well-known $REMn_6Sn_6$ and $AM_3X_5$ families[12,25]. Due to the vdW interactions between layers and the unique stacking pattern of the three elements, this material system can be considered to exhibit quasi-2D characteristic with hexagonal symmetry. In particular, it is rather rare for the edge-sharing nickel-chalcogen square planes to form a hexagonal prism to the best of our knowledge. In most cases, metal-chalcogen squares would prefer to form an infinite 2D square lattice, like the Ni-O planes in the superconductor $Nd_{0.8}Sr_{0.2}NiO_2$[49]. The uniqueness of this feature might provide a platform for investigating novel physical properties, such as the interplay of the kagome, triangular and honeycomb lattices. Also, due to the typically large ionic radii of alkali metals, $Rb_2Ni_3S_4$ (or $Cs_2Ni_3Se_4$) exhibits a large interlayer spacing (~10 Å). Together with its vdW nature, the obtained crystals can be easily cleaved along *ac* planes.

Figure 2(a) presents the powder diffraction pattern and the corresponding Rietveld refinement results of the as-grown $Rb_2Ni_3S_4$ polycrystalline sample. The obtained lattice constant values, $a = 9.9262(2)$ Å, $b = 13.6490(3)$ Å, $c = 5.8755(9)$ Å, are consistent with previous literature[46]. The $R_{wp}$ factors after the final iteration of the Rietveld refinement is 6.263 %, reflecting a reliable fitting result. Figure 2(b) and Figure 2(c) show the powder XRD patterns taken on the freshly exfoliated, platelet-like $Rb_2Ni_3S_4$ and $Cs_2Ni_3Se_4$ single crystal samples, respectively. Only (0 *k* 0) peaks can be detected, indicating the single-crystalline nature of both crystals with the *b*-axis perpendicular to the exposed surface. In the upcoming experiments, our primary focus will be on studying the properties under a magnetic field applied parallel or perpendicular to the sample surface, namely, perpendicular or parallel to the *b*-axis.

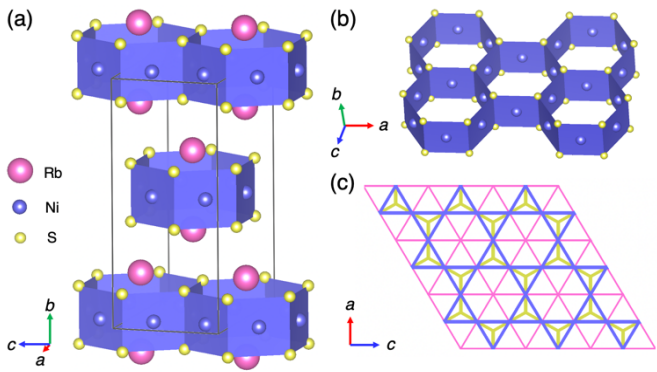

**Figure 1.** (a) Schematic of the crystal structure of $Rb_2Ni_3S_4$. (b) The edge-sharing Ni-S square planes form hexagonal prisms arranged along the *ac* plane. (c) Stacking order of transition-metal kagome (blue), sulfur honeycomb (yellow) and alkali-metal triangular lattices (pink), viewing along *b*-axis.

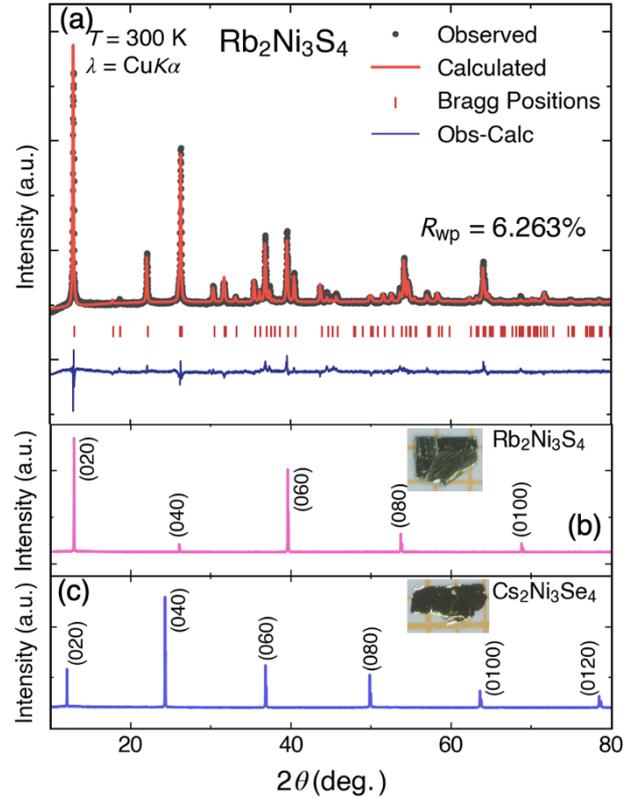

**Figure 2.** (a) X-ray diffraction (XRD) pattern along with Rietveld refinement results of $Rb_2Ni_3S_4$ polycrystalline sample. (b, c) XRD patterns of $Rb_2Ni_3S_4$ (b) and $Cs_2Ni_3Se_4$ (c). Insets show photographs of single crystals on a 1-mm grid scale.

Figure 3(a) shows the temperature dependence of the magnetic susceptibility $\chi(T) = M/H$ for the $Rb_2Ni_3S_4$ and $Cs_2Ni_3Se_4$ single crystals under a magnetic field of 0.5 T, applied along and perpendicular to the *b* axis, respectively. The weak temperature-dependent $\chi(T)$ over the measured temperature range indicates a nonmagnetic behavior with a very small net magnetic moment of ~0.001 $\mu_B$ per Ni atom, which is also consistent with the field-dependent magnetization in Figure S3 (see Supplementary Information). Since the $3d^8$ electrons of $Ni^{2+}$ are squarely coordinated by chalcogenide anions, the local square-planar environment leads to a low spin ($S = 0$) configuration, as illustrated in Figure 3(b), and consequently no magnetism or geometric frustration effect is expected. Still, there is a small but apparent anisotropic behavior in the susceptibility between the *b*-axis and the *ac* plane, but it seems quite common in layered quasi-2D materials[22]. Despite the previous inconsistent opinions about its magnetic properties[40,42,43], our results favor the nonmagnetic ground state while the weak ferromagnetism observed in early studies[42] might be attributed to the degraded samples in air or humidity.

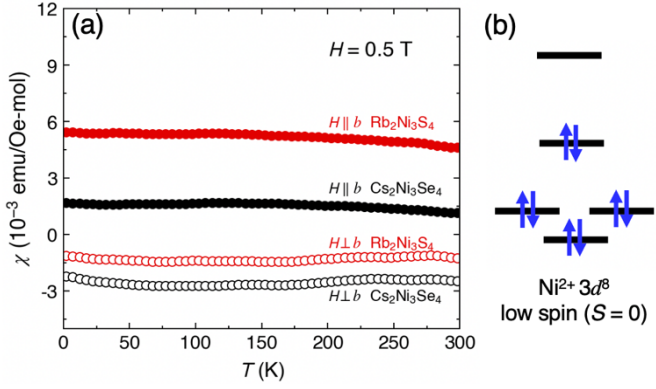

**Figure 3.** Magnetic properties of $Rb_2Ni_3S_4$ and $Cs_2Ni_3Se_4$. (a) Temperature dependence of magnetic susceptibility under a field of 0.5 T applied along (solid circles) and perpendicular to (open circles) *b* axis. (b) Spin configuration of $Ni^{2+}$ in the square planar configuration.

Temperature dependence of the resistivity was investigated on bulk $Rb_2Ni_3S_4$ and $Cs_2Ni_3Se_4$ single crystals (~ 5 × 3 × 0.05 $mm^3$), as depicted in Figure 4(a). The crystals display an average resistivity of 82 ± 6 Ω·cm and 1.2 ± 2 Ω·cm at room temperature, respectively. Both $Rb_2Ni_3S_4$ and $Cs_2Ni_3Se_4$ show semiconducting behavior with the resistivity increasing sharply as temperature decreases. Within a temperature range from 250 K to 150 K, the resistivity behavior can be well described by a thermally activated model, $\rho = \rho_0 \exp(-\Delta E/2kT)$, with the activation energy ($\Delta E$) equal to 0.37 eV for $Rb_2Ni_3S_4$ and 0.05 eV for $Cs_2Ni_3Se_4$, as shown in the inset of Figure 4(a). The fitted values are slightly smaller than those reported by previous band structure calculations[44], which is probably due to the interlayer alkali metal occupancy issue.

Notably, a kink in resistivity appears near 260 K for $Rb_2Ni_3S_4$ samples, as shown in Figure 4(b). This resistivity kink has been observed in all measured $Rb_2Ni_3S_4$ samples at the same temperature (see Supplementary Information, Figure S5), indicating an intrinsic characteristic of $Rb_2Ni_3S_4$. However, no such behavior was observed in $Cs_2Ni_3Se_4$. Considering that there is no structural transition or distortion observed from previous Raman spectroscopy[41], nor magnetic transition evidence in the magnetic susceptibility near 260 K, this phase transition is most likely of electronic origin, as reported in some other kagome materials[21,35]. Further research is needed to elucidate the exact origin of such an anomaly. Representative Hall resistivity data for $Rb_2Ni_3S_4$ and $Cs_2Ni_3Se_4$ measured at 240 K are shown in Figure 4(c), suggesting that our synthesized $Rb_2Ni_3S_4$ and $Cs_2Ni_3Se_4$ single crystal samples are hole-doped. This feature may originate from the absence of interlayer alkali metals as the alkali metals are relatively labile in layered quasi-2D structure, for example in the superconductor $K_xFe_2Se_2$ ($0 \leq x \leq 1.0$)[50]. Thus this feature offers the possibility of alkali metal intercalation or deintercalation by chemical methods[51–53].

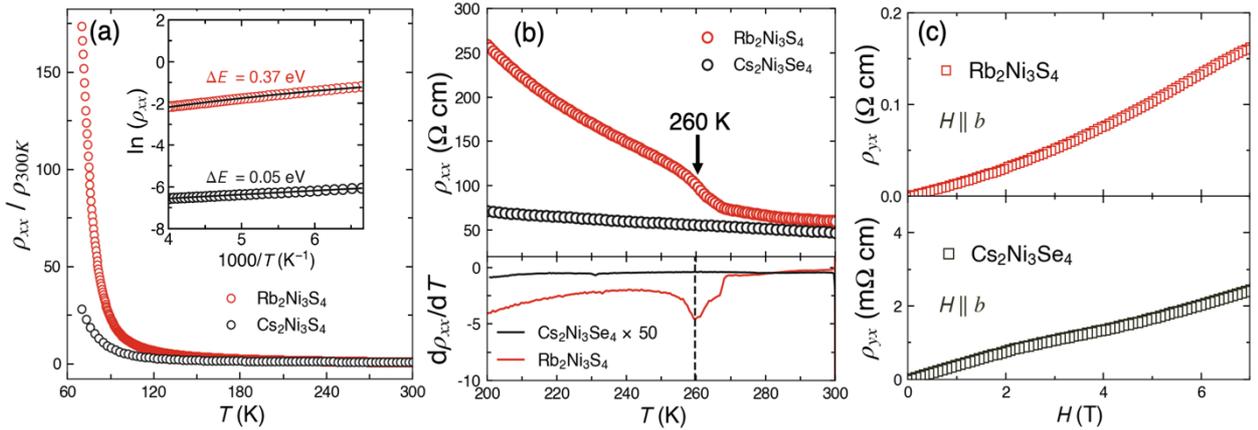

**Figure 4.** Electrical transport properties of $Rb_2Ni_3S_4$ and $Cs_2Ni_3Se_4$. (a) Resistivity as a function of temperature $\rho_{xx}(T)$ of $Rb_2Ni_3S_4$ and $Cs_2Ni_3Se_4$ measured at zero field. The data have been normalized to room temperature for better comparison. Inset shows the log $\rho_{xx}$ versus 1000/T between 150-250 K. The black lines represent an Arrhenius fitting. (b) A zoomed-in figure of (a) for 220 K < T < 300 K and the corresponding derivative curves. The resistivity of $Cs_2Ni_3Se_4$ has been enlarged by 50 times for comparison. (c) Hall resistivity $\rho_{yx}$ measured at 240 K.

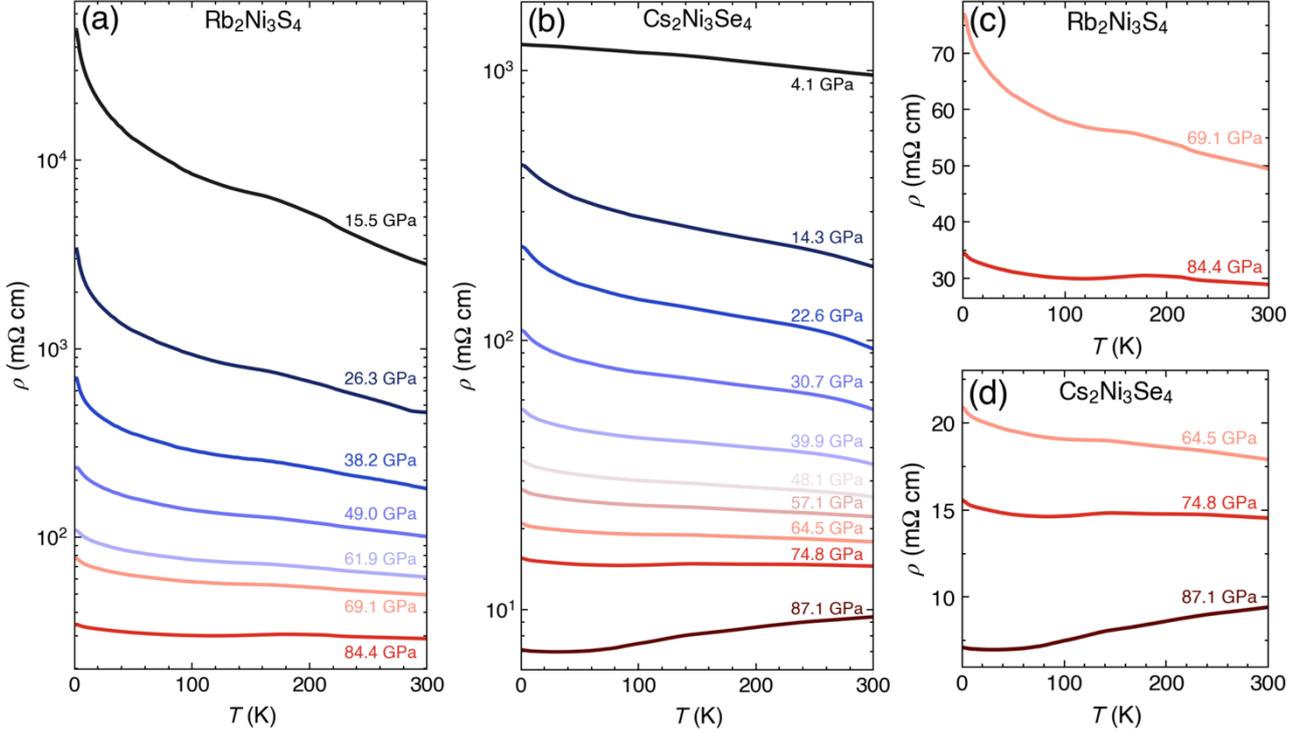

**Figure 5.** (a) Electrical resistivity of $Rb_2Ni_3S_4$ as a function of temperature for pressures up to 84.4 GPa. (b) Electrical resistivity of $Cs_2Ni_3Se_4$ as a function of temperature for pressures up to 87.1 GPa. (c-d) The $\rho(T)$ curves in the vicinity of the metallization.

Figure 5 shows the temperature dependence of resistivity for $Rb_2Ni_3S_4$ and $Cs_2Ni_3Se_4$ single crystals in the pressure range of 4.1 - 87.1 GPa. As shown in Figure 5(a) and 5(b), the $\rho(T)$ curve exhibits negative $d\rho/dT$ throughout all temperatures below 70 GPa, indicative of a typical semiconducting behavior. Interestingly, positive $d\rho/dT$ is observed in the intermediate temperature region for $Rb_2Ni_3S_4$ at 84.4 GPa and $Cs_2Ni_3Se_4$ at 74.8 GPa, as shown in Figure 5(c) and 5(d). This suggests the presence of metallization behavior close to the IMT boundary. Furthermore, at 87.1 GPa, the resistivity of $Cs_2Ni_3Se_4$ initially decreases upon cooling from 300 K and reaches a plateau at 30 K. Thus, $Cs_2Ni_3Se_4$ ultimately undergoes a metallization under pressure of 87.1 GPa, while $Rb_2Ni_3S_4$ shows the tendency to be metalized. High-pressure Raman spectroscopy experiments carried out on $Cs_2Ni_3Se_4$, as shown in Figure 6, reveal two sharp Raman modes at 180 and 230 cm$^{-1}$ at 7.1 GPa. As pressure increases, the two vibration modes exhibit blue shift monotonically and become weaker in intensity, as also depicted in Figure 6(b). No anomaly is observed in the Raman shift as a function of pressure up to the highest achievable pressure (85.1 GPa) (Figure 6(b)), implying the absence of a structural phase transition. However, more definitive evidence from high-pressure XRD is needed. Therefore, the pressure-induced IMT of $Cs_2Ni_3Se_4$ could be attributed to the bandgap closure under compression.

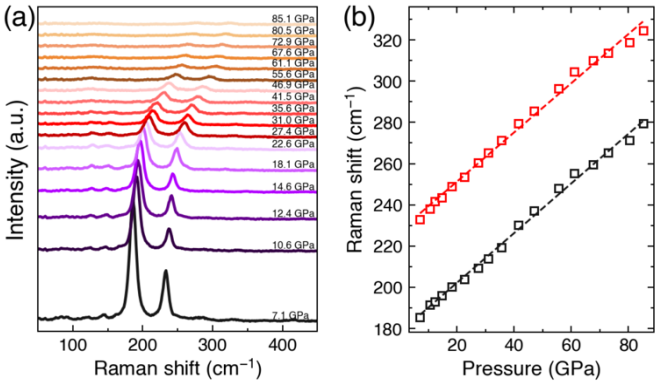

**Figure 6.** (a) Raman spectra at various pressures for $Cs_2Ni_3Se_4$. (b) Raman mode frequencies for $Cs_2Ni_3Se_4$ under compression. The dashed lines represent linear fittings.

## 4. Conclusions

In conclusion, $A_2M_3X_4$ compounds synthesized by our methods are vdW nonmagnetic, hole-doped semiconductors featuring a stacked transition-metal-based kagome lattice. The nonmagnetic behavior in $Rb_2Ni_3S_4$ and $Cs_2Ni_3Se_4$ is closely related to the square-planar environment of $Ni^{2+}$, which results in the low spin state of $S = 0$ for $Ni^{2+}$. The hole-doped semiconducting characteristic is likely originated from the vacancies of alkali metals in between the vdW layers. Since alkali metals can be readily intercalated or de-intercalated from layered chalcogenides, the nonmagnetic and semiconducting properties can be easily modified by either electron or hole doping, leading to magnetism on geometric frustrated lattices or metallic behavior with topological flat bands. A kink in transport measurement around 260 K is observed in $Rb_2Ni_3S_4$, which implies a possible electronic structure transition and leaves a focal point for future studies. In addition, we notice that $Cs_2Ni_3Se_4$ shows an IMT in transport measurements under pressure up to 87.1 GPa, probably due to a bandgap closure under compression, while $Rb_2Ni_3S_4$ shows the tendency to be metalized. Based on this work, we find the studied vdW kagome chalcogenide $A_2M_3X_4$ is a highly tunable material system with rich structural features, and it hosts the potential to be tuned towards geometric frustrated magnetism, topological electronics and superconductivity for future investigations. Our findings attest to the close relationship between the crystal structure and physical properties of layered kagome $A_2M_3X_4$, providing further insights into their potential for new material design.

**Acknowledgment**

The work was supported by the Ministry of Science and Technology of China under 2022YFA1402702, National Natural Science Foundation of China with Grants No. 12334008, and No. 12374148. R.Z. also acknowledges the support for the project by the National Key R&D of China under 2021YFA1401600. Y.P.Q. thanks the support from the Natural Science Foundation of China (Grant No. 52272265), the National Key R&D Program of China (Grant Nos. 2018YFA0704300, 2023YFA1607400), and Shanghai Science and Technology Plan (Grant No. 21DZ2260400).

**References**

[1] Yin J-X, Lian B and Hasan M Z 2022 Topological kagome magnets and superconductors *Nature* **612** 647–57

[2] Wang Y, Wu H, McCandless G T, Chan J Y and Ali M N 2023 Quantum states and intertwining phases in kagome materials *Nat Rev Phys* **5** 635–58

[3] Norman M R 2016 *Colloquium*: Herbertsmithite and the search for the quantum spin liquid *Rev. Mod. Phys.* **88** 041002

[4] Kiesel M L and Thomale R 2012 Sublattice interference in the kagome Hubbard model *Phys. Rev. B* **86** 121105

[5] Kiesel M L, Platt C and Thomale R 2013 Unconventional Fermi Surface Instabilities in the Kagome Hubbard Model *Phys. Rev. Lett.* **110** 126405

[6] Liu E, Sun Y, Kumar N, Muechler L, Sun A, Jiao L, Yang S-Y, Liu D, Liang A, Xu Q, Kroder J, Süß V, Borrmann H, Shekhar C, Wang Z, Xi C, Wang W, Schnelle W, Wirth S, Chen Y, Goennenwein S T B and Felser C 2018 Giant anomalous Hall effect in a ferromagnetic kagome-lattice semimetal *Nature Phys* **14** 1125–31

[7] Morali N, Batabyal R, Nag P K, Liu E, Xu Q, Sun Y, Yan B, Felser C, Avraham N and Beidenkopf H 2019 Fermi-arc diversity on surface terminations of the magnetic Weyl semimetal $Co_3Sn_2S_2$ *Science* **365** 1286–91

[8] Liu D F, Liang A J, Liu E K, Xu Q N, Li Y W, Chen C, Pei D, Shi W J, Mo S K, Dudin P, Kim T, Cacho C, Li G, Sun Y, Yang L X, Liu Z K, Parkin S S P, Felser C and Chen Y L 2019 Magnetic Weyl semimetal phase in a Kagomé crystal *Science* **365** 1282–5

[9] Tang E, Mei J-W and Wen X-G 2011 High-Temperature Fractional Quantum Hall States *Phys. Rev. Lett.* **106** 236802

[10] Ko W-H, Lee P A and Wen X-G 2009 Doped kagome system as exotic superconductor *Phys. Rev. B* **79** 214502

[11] Fu M, Imai T, Han T-H and Lee Y S 2015 Evidence for a gapped spin-liquid ground state in a kagome Heisenberg antiferromagnet *Science* **350** 655–8

[12] Venturini G, Idrissi B C E and Malaman B 1991 Magnetic properties of $RMn_6Sn_6$ (R = Sc, Y, Gd−Tm, Lu) compounds with $HfFe_6Ge_6$ type structure *Journal of Magnetism and Magnetic Materials* **94** 35–42

[13] Yin J-X, Ma W, Cochran T A, Xu X, Zhang S S, Tien H-J, Shumiya N, Cheng G, Jiang K, Lian B, Song Z, Chang G, Belopolski I, Multer D, Litskevich M, Cheng Z-J, Yang X P, Swidler B, Zhou H, Lin H, Neupert T, Wang Z, Yao N, Chang T-R, Jia S and Zahid Hasan M 2020

Quantum-limit Chern topological magnetism in TbMn$_6$Sn$_6$ *Nature* **583** 533–6

[14] Wang Q, Xu Y, Lou R, Liu Z, Li M, Huang Y, Shen D, Weng H, Wang S and Lei H 2018 Large intrinsic anomalous Hall effect in half-metallic ferromagnet Co$_3$Sn$_2$S$_2$ with magnetic Weyl fermions *Nat Commun* **9** 3681

[15] Liu Z, Li M, Wang Q, Wang G, Wen C, Jiang K, Lu X, Yan S, Huang Y, Shen D, Yin J-X, Wang Z, Yin Z, Lei H and Wang S 2020 Orbital-selective Dirac fermions and extremely flat bands in frustrated kagome-lattice metal CoSn *Nat Commun* **11** 4002

[16] Lin Z, Choi J-H, Zhang Q, Qin W, Yi S, Wang P, Li L, Wang Y, Zhang H, Sun Z, Wei L, Zhang S, Guo T, Lu Q, Cho J-H, Zeng C and Zhang Z 2018 Flatbands and Emergent Ferromagnetic Ordering in Fe$_3$Sn$_2$ Kagome Lattices *Phys. Rev. Lett.* **121** 096401

[17] Ye L, Kang M, Liu J, Von Cube F, Wicker C R, Suzuki T, Jozwiak C, Bostwick A, Rotenberg E, Bell D C, Fu L, Comin R and Checkelsky J G 2018 Massive Dirac fermions in a ferromagnetic kagome metal *Nature* **555** 638–42

[18] Li Y, Wang Q, DeBeer-Schmitt L, Guguchia Z, Desautels R D, Yin J-X, Du Q, Ren W, Zhao X, Zhang Z, Zaliznyak I A, Petrovic C, Yin W, Hasan M Z, Lei H and Tranquada J M 2019 Magnetic-Field Control of Topological Electronic Response near Room Temperature in Correlated Kagome Magnets *Phys. Rev. Lett.* **123** 196604

[19] Kang M, Ye L, Fang S, You J-S, Levitan A, Han M, Facio J I, Jozwiak C, Bostwick A, Rotenberg E, Chan M K, McDonald R D, Graf D, Kaznatcheev K, Vescovo E, Bell D C, Kaxiras E, Van Den Brink J, Richter M, Prasad Ghimire M, Checkelsky J G and Comin R 2020 Dirac fermions and flat bands in the ideal kagome metal FeSn *Nat. Mater.* **19** 163–9

[20] Teng X, Chen L, Ye F, Rosenberg E, Liu Z, Yin J-X, Jiang Y-X, Oh J S, Hasan M Z, Neubauer K J, Gao B, Xie Y, Hashimoto M, Lu D, Jozwiak C, Bostwick A, Rotenberg E, Birgeneau R J, Chu J-H, Yi M and Dai P 2022 Discovery of charge density wave in a kagome lattice antiferromagnet *Nature* **609** 490–5

[21] Teng X, Oh J S, Tan H, Chen L, Huang J, Gao B, Yin J-X, Chu J-H, Hashimoto M, Lu D, Jozwiak C, Bostwick A, Rotenberg E, Granroth G E, Yan B, Birgeneau R J, Dai P and Yi M 2023 Magnetism and charge density wave order in kagome FeGe *Nat. Phys.* **19** 814–22

[22] Park S, Kang S, Kim H, Lee K H, Kim P, Sim S, Lee N, Karuppannan B, Kim J, Kim J, Sim K I, Coak M J, Noda Y, Park C-H, Kim J H and Park J-G 2020 Kagome van-der-Waals Pd$_3$P$_2$S$_8$ with flat band *Sci Rep* **10** 20998

[23] Wang Q, Qiu X-L, Pei C, Gong B-C, Gao L, Zhao Y, Cao W, Li C, Zhu S, Zhang M, Chen Y, Liu K and Qi Y 2023 Superconductivity emerging from a pressurized van der Waals kagome material Pd$_3$P$_2$S$_8$ *New J. Phys.* **25** 043001

[24] Yang T Y, Wan Q, Song J P, Du Z, Tang J, Wang Z W, Plumb N C, Radovic M, Wang G W, Wang G Y, Sun Z, Yin J-X, Chen Z H, Huang Y B, Yu R, Shi M, Xiong Y M and Xu N 2022 Fermi-level flat band in a kagome magnet *Quantum Front* **1** 14

[25] Ortiz B R, Gomes L C, Morey J R, Winiarski M, Bordelon M, Mangum J S, Oswald I W H, Rodriguez-Rivera J A, Neilson J R, Wilson S D, Ertekin E, McQueen T M and Toberer E S 2019 New kagome prototype materials: discovery of KV$_3$Sb$_5$, RbV$_3$Sb$_5$, and CsV$_3$Sb$_5$ *Phys. Rev. Materials* **3** 094407

[26] Chen K Y, Wang N N, Yin Q W, Gu Y H, Jiang K, Tu Z J, Gong C S, Uwatoko Y, Sun J P, Lei H C, Hu J P and Cheng J-G 2021 Double Superconducting Dome and Triple Enhancement of T c in the Kagome Superconductor CsV$_3$Sb$_5$ under High Pressure *Phys. Rev. Lett.* **126** 247001

[27] Ortiz B R, Sarte P M, Kenney E M, Graf M J, Teicher S M L, Seshadri R and Wilson S D 2021 Superconductivity in the Z$_2$ kagome metal KV$_3$Sb$_5$ *Phys. Rev. Materials* **5** 034801

[28] Jiang Y-X, Yin J-X, Denner M M, Shumiya N, Ortiz B R, Xu G, Guguchia Z, He J, Hossain M S, Liu X, Ruff J, Kautzsch L, Zhang S S, Chang G, Belopolski I, Zhang Q, Cochran T A, Multer D, Litskevich M, Cheng Z-J, Yang X P, Wang Z, Thomale R, Neupert T, Wilson S D and Hasan M Z 2021 Unconventional chiral charge order in

kagome superconductor KV$_3$Sb$_5$ *Nat. Mater.* **20** 1353–7

[29] Wang Q, Kong P, Shi W, Pei C, Wen C, Gao L, Zhao Y, Yin Q, Wu Y, Li G, Lei H, Li J, Chen Y, Yan S and Qi Y 2021 Charge Density Wave Orders and Enhanced Superconductivity under Pressure in the Kagome Metal CsV$_3$Sb$_5$ *Advanced Materials* **33** 2102813

[30] Chen H, Yang H, Hu B, Zhao Z, Yuan J, Xing Y, Qian G, Huang Z, Li G, Ye Y, Ma S, Ni S, Zhang H, Yin Q, Gong C, Tu Z, Lei H, Tan H, Zhou S, Shen C, Dong X, Yan B, Wang Z and Gao H-J 2021 Roton pair density wave in a strong-coupling kagome superconductor *Nature* **599** 222–8

[31] Nie L, Sun K, Ma W, Song D, Zheng L, Liang Z, Wu P, Yu F, Li J, Shan M, Zhao D, Li S, Kang B, Wu Z, Zhou Y, Liu K, Xiang Z, Ying J, Wang Z, Wu T and Chen X 2022 Charge-density-wave-driven electronic nematicity in a kagome superconductor *Nature* **604** 59–64

[32] Mielke C, Das D, Yin J-X, Liu H, Gupta R, Jiang Y-X, Medarde M, Wu X, Lei H C, Chang J, Dai P, Si Q, Miao H, Thomale R, Neupert T, Shi Y, Khasanov R, Hasan M Z, Luetkens H and Guguchia Z 2022 Time-reversal symmetry-breaking charge order in a kagome superconductor *Nature* **602** 245–50

[33] Neupert T, Denner M M, Yin J-X, Thomale R and Hasan M Z 2022 Charge order and superconductivity in kagome materials *Nat. Phys.* **18** 137–43

[34] Hu Y, Le C, Zhang Y, Zhao Z, Liu J, Ma J, Plumb N C, Radovic M, Chen H, Schnyder A P, Wu X, Dong X, Hu J, Yang H, Gao H-J and Shi M 2023 Non-trivial band topology and orbital-selective electronic nematicity in a titanium-based kagome superconductor *Nat. Phys.* **19** 1827–33

[35] Li H, Cheng S, Ortiz B R, Tan H, Werhahn D, Zeng K, Johrendt D, Yan B, Wang Z, Wilson S D and Zeljkovic I 2023 Electronic nematicity without charge density waves in titanium-based kagome metal *Nat. Phys.* **19** 1591–8

[36] Zhao H, Li H, Ortiz B R, Teicher S M L, Park T, Ye M, Wang Z, Balents L, Wilson S D and Zeljkovic I 2021 Cascade of correlated electron states in the kagome superconductor CsV$_3$Sb$_3$ *Nature* **599** 216–21

[37] Ortiz B R, Teicher S M L, Hu Y, Zuo J L, Sarte P M, Schueller E C, Abeykoon A M M, Krogstad M J, Rosenkranz S, Osborn R, Seshadri R, Balents L, He J and Wilson S D 2020 CsV$_3$Sb$_5$: A Z$_2$ Topological Kagome Metal with a Superconducting Ground State *Phys. Rev. Lett.* **125** 247002

[38] Liang Z, Hou X, Zhang F, Ma W, Wu P, Zhang Z, Yu F, Ying J-J, Jiang K, Shan L, Wang Z and Chen X-H 2021 Three-Dimensional Charge Density Wave and Surface-Dependent Vortex-Core States in a Kagome Superconductor CsV$_3$Sb$_5$ *Phys. Rev. X* **11** 031026

[39] Bronger W, Eyck J, Rüdorff W and Stöussel A 1970 Über Thio- und Selenoniccolate und -palladate der schweren Alkalimetalle *Zeitschrift anorg allge chemie* **375** 1–7

[40] Elder S H, Jobic S, Brec R, Gelabert M and DiSalvo F J 1996 Structural and electronic properties of K$_2$Ni$_3$S$_4$, a pseudo-two dimensional compound with a honeycomb-like arrangement *Journal of Alloys and Compounds* **235** 135–42

[41] Hasegawa T, Inui M, Hondou K, Fujiwara Y, Kato T and Iio K 2004 Raman spectroscopy on ternary transition metal chalcogenide Rb$_2$Ni$_3$S$_4$ *Journal of Alloys and Compounds* **364** 199–207

[42] Kato T, Hondou K and Iio K 1998 Magnetism of the Ni$^{2+}$ kagome lattice in Rb$_2$Ni$_3$S$_4$ *Journal of Magnetism and Magnetic Materials* **177–181** 591–2

[43] Fukamachi T, Kobayashi Y, Nakamura A, Harashina H and Sato M 1999 Magnetization Measurements and $^{87}$Rb-NMR on Rb$_2$M$_3$S$_4$ (M = Ni and Pd) with the Kagome Lattice *Journal of the Physical Society of Japan* **68** 3668–72

[44] Nawai S, Okazaki K, Mizokawa T, Fujimori A, Hondou K, Fujiwara Y, Iio K, Usuda M and Hamada N 2004 Electronic structure of the Kagomé lattice compound Rb$_2$Ni$_3$S$_4$ *Phys. Rev. B* **69** 045103

[45] Li Q, Wu Y, Fan X, Zhang Y-J, Zhu X, Zhu Z, Li Y and Wen H-H 2022 Superconductivity arising from pressure-induced emergence of a Fermi surface in the kagome-lattice chalcogenide $Rb_2Pd_3Se_4$ *Phys. Rev. B* **106** 214501

[46] Bronger W, Rennau R and Schmitz D 1991 Schichtstrukturen ternärer Chalkogenide $A_2M_3X_4$ (A ≙ K, Rb, Cs; M ≙ Ni, Pd, Pt; X ≙ S, Se) *Zeitschrift anorg allge chemie* **597** 27–32

[47] Rietveld H M 1969 A profile refinement method for nuclear and magnetic structures *J Appl Crystallogr* **2** 65–71

[48] Pei C, Zhang J, Wang Q, Zhao Y, Gao L, Gong C, Tian S, Luo R, Li M, Yang W, Lu Z-Y, Lei H, Liu K and Qi Y 2023 Pressure-induced superconductivity at 32 K in $MoB_2$ *National Science Review* **10** nwad034

[49] Li D, Lee K, Wang B Y, Osada M, Crossley S, Lee H R, Cui Y, Hikita Y and Hwang H Y 2019 Superconductivity in an infinite-layer nickelate *Nature* **572** 624–7

[50] Guo J, Jin S, Wang G, Wang S, Zhu K, Zhou T, He M and Chen X 2010 Superconductivity in the iron selenide $K_xFe_2Se_2$ ($0 \leq x \leq 1.0$) *Phys. Rev. B* **82** 180520

[51] Murphy D W, Carides J N, Di Salvo F J, Cros C and Waszczak J V 1977 Cathodes for nonaqueous lithium batteries based on $VS_2$ *Materials Research Bulletin* **12** 825–30

[52] Van Bruggen C F, Haange R J, Wiegers G A and De Boer D K G 1980 $CrSe_2$, a new layered dichalcogenide *Physica B+C* **99** 166–72

[53] Song X, Schneider S N, Cheng G, Khoury J F, Jovanovic M, Yao N and Schoop L M 2021 Kinetics and Evolution of Magnetism in Soft-Chemical Synthesis of $CrSe_2$ from $KCrSe_2$ *Chem. Mater.* **33** 8070–8